\documentclass[conference]{IEEEtran}
\IEEEoverridecommandlockouts
\usepackage{stfloats,color}
\usepackage{amsfonts}
\usepackage[cmex10]{amsmath}
\usepackage{graphicx}
\usepackage{setspace}
\usepackage{cite}
\usepackage{array}
\usepackage{algorithmic}
\usepackage{mdwmath}
\usepackage{mdwtab}
\usepackage[center]{caption}
\usepackage{fixltx2e}
\usepackage{url}
\usepackage{times}
\usepackage{epsfig}
\usepackage{latexsym}
\usepackage{epstopdf}
\usepackage{verbatim}
\usepackage{units}
\usepackage{amsthm}
\usepackage{mdwlist}

\allowdisplaybreaks
\usepackage{soul}\graphicspath{{./figures/}}

\begin{document}

\sloppy

\title{Coverage Probability Analysis for Wireless Networks Using Repulsive Point Processes}

\author{\IEEEauthorblockN{ Abdelrahman M. Ibrahim, Tamer ElBatt$^2$ , Amr El-Keyi}
\IEEEauthorblockA{Wireless Intelligent Networks Center (WINC), Nile University, Giza, Egypt. \\  
Email: abdelrahman.ibrahim@nileu.edu.eg  ,
\{telbatt, aelkeyi\}@nileuniversity.edu.eg}}
\maketitle

\begin{abstract}
The recent witnessed evolution of cellular networks from a carefully planned deployment to more irregular, heterogeneous deployments of Macro, Pico and Femto-BSs motivates new analysis and design approaches.
In this paper, we analyze the coverage probability in cellular networks assuming repulsive point processes for the base station deployment. In particular, we characterize, analytically using stochastic geometry, the downlink probability of coverage under a    Matern hardcore point process to ensure minimum distance between the randomly located base stations. Assuming a mobile user connects to the nearest base station and Rayleigh fading, we derive two lower bounds expressions on the downlink probability of coverage that is within $4\%$ from the simulated scenario. To validate our model, we compare the probability of coverage of the Matern hardcore topology against an actual base station deployment obtained from a public database. The comparison shows that the actual base station deployment can be fitted by setting the appropriate Matern point process density.
\end{abstract}

\footnotetext[1]{This publication was made possible by NPRP grant $\#$ 5-782-2-322 from the Qatar National Research Fund (a member of Qatar Foundation). The statements made herein are solely the responsibility of the authors.}
\footnotetext[2]{Tamer ElBatt is also affiliated with the EECE Dept., Faculty of Engineering, Cairo University.}

\section{Introduction}
\begin{figure}
\includegraphics[width=0.40\textwidth, height=0.23\textheight]{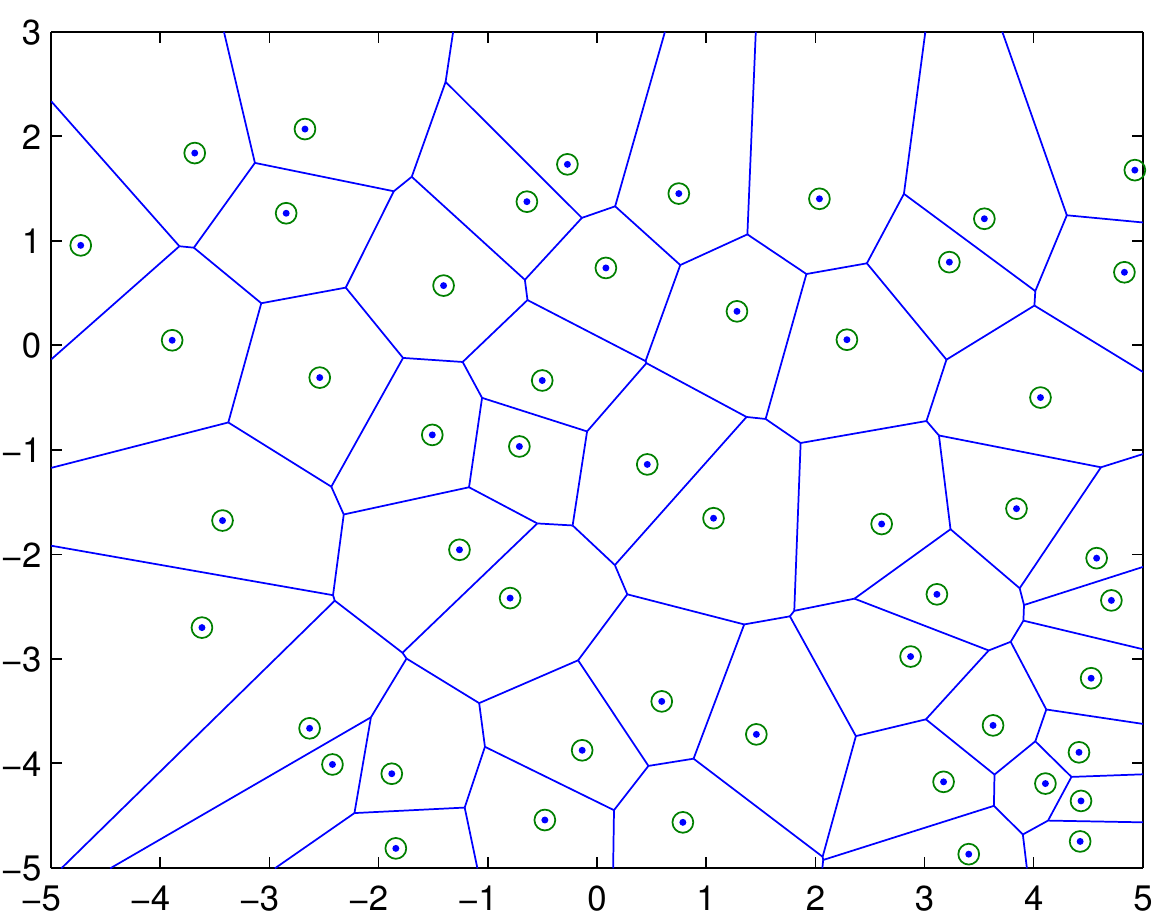}
\centering
\caption{Actual BS deployment from a rural area \cite{sitefinder}}\label{fig:actual}
\end{figure}

Cellular networks capacity is fundamentally limited by the intensity of the received power and interference. Both are highly dependent on the spatial locations of the base stations (BSs). 
By far, the most popular approach used in modeling the BSs topology is the hexagonal grid model adopted by standard bodies such as the 3rd Generation Partnership Project (3GPP). Grid models 
are highly idealized models which do not accurately capture the actual BSs topology. In reality, cells radii differ from one cell to another due to differences in the transmitted powers and the user density as shown for a real deployment in Fig. \ref{fig:actual}.\\  
\indent The most common information theoretic downlink model for cellular networks is the Wyner Model \cite{wyner1994shannon} due to its mathematical tractability. However, it is a simplified one dimensional model that sets the Signal-to-Interference ratio (SIR) as a constant. Moreover,  the Wyner Model is impractical for OFDMA systems where the SIR values vary dramatically across the cell \cite{xu2011accuracy}. Also, the Wyner model fixes the user location, therefore it is highly inaccurate for analyzing the probability of coverage ($P_{c}$).\\   
\indent The recent witnessed evolution of cellular networks from a carefully planned deployment to more irregular, heterogeneous deployments of Macro, Pico and Femto-BSs renders the hexagonal and regular deployment models of limited utility. This, in turn, motivates recent studies, tools and results \cite{andrews2010primer} \cite{dhillon2012modeling} \cite{haenggi2009interference} inspired by stochastic geometry. A prominent approach is to use random spatial models from stochastic geometry \cite{baddeley2007stochastic} \cite{serfozo2009basics} to capture the real deployment as accurately as possible. Stochastic geometry allows us to study the average behavior over many spatial realizations of a network where the nodes locations are derived from a point process (PP) \cite{andrews2010primer}  \cite{haenggi2009interference} \cite{haenggi2009stochastic}. Most of the stochastic geometry work on cellular networks focus on the case where the BS deployment follows a Poisson point process (PPP). In \cite{andrews2011tractable}, the points derived from a PPP are independent which significantly simplifies the analysis. However, this is far from reality since the BSs locations in real cellular networks are not totally independent. Instead, they are planned deployments with a degree of randomness due to irregular terrains and hot-spots as shown in \cite{andrews2011tractable} and \cite{taylor2012pairwise}. 
\vspace{-0.04in}
\subsection{Scope}
In this work, we extend the coverage analysis of a PPP by using a stationary point process that captures the repulsion between BSs. We generalize the independent PPP analytical framework in \cite{andrews2011tractable} to a Matern hardcore (MHC) point process \cite{stoyan1985one} which maintains a minimum separation between BSs in an attempt to capture real deployments.
\vspace{-0.04in}
\subsection{Related Work}

The recent work in \cite{andrews2011tractable} introduced a stochastic geometry framework for the analysis of coverage and rate in 1-tier cellular networks. In this framework, Macro-BSs locations follow a homogeneous PPP and the users locations are derived from an independent PPP. Also, the users are assumed to connect to the nearest BS. The authors derived closed form expressions for the probability of coverage under Rayleigh fading. Also, they compared the $P_c$ of the PPP model and the grid model against an actual data from a real BS deployment. The $P_c$ comparison showed that the PPP model can be considered a lower bound to the real deployment and the grid model can be considered an upper bound.\\ 
\indent Recent studies to extend the PPP framework to non-Poisson point processes, in order to model the dependence between BSs in cellular networks, can be found in \cite{taylor2012pairwise} \cite{busson2009interference}. One of the main difficulties in the analysis of non-Poisson point processes is the mathematical intractability attributed to the absence of a closed form expression for the probability generating functional (PGFL) of the underlying node distribution. An alternative approach is presented in \cite{ganti2010new} to overcome the PGFL hurdle by using Weierstrass inequality \cite{klamkin1970extensions}. The authors derived bounds on the probability of coverage utilizing the second order density of the underlying node distribution. However, these bounds are suggested for very small densities less than $0.04$ and diverge as the SINR threshold increases. Therefore, we overcome the shortcomings of this approach by proposing a general framework based on the Matern hardcore point process \cite{stoyan1985one} in order to find a lower bound on the probability of coverage.
\vspace{-0.04in}
\subsection{Contributions}
Our contribution in this paper is multi-fold. First, we extend the stochastic geometry framework presented in \cite{andrews2011tractable} to model the BSs locations using a MHC point process, incorporating dependence between deployment points as encountered in practice. Second, we derive the MHC empty space distribution. Third, we overcome the PGFL hurdle by applying Jensen's inequality and the inequality proposed in Conjecture 1 to establish tight lower bounds on the MHC coverage probability. 
Finally, we compare the coverage probability of PPP, square grid and MHC deployments to an actual BS deployment from a rural area \cite{sitefinder}. We also compare the simulated MHC coverage probability to the analytical lower bounds which confirm their  tightness, especially for the Conjecture 1 inequality-based bound.\\ 
\indent The rest of this paper is organized as follows. In Section \ref{sec:background}, we present a background on the stochastic geometry tools and the point processes used in this paper. Afterwards, we present the system model in Section \ref{sec:systemmodel}. In Section \ref{sec:coverage}, we present our main analytical results and establish lower bounds on the coverage probability. In Section \ref{sec:results}, we provide numerical results to support our analytical findings. Finally, conclusions are drawn and potential directions for future research are pointed out in Section \ref{sec:conclusion}.

\section{Background: Stochastic Geometry}\label{sec:background}
\subsection{Spatial Point Processes}

A spatial point process (PP) $\Phi $ is a random collection of points in space. A PP is simple if no two points are at the same location, i.e.\ $x \neq y$  for any $x,y \in \Phi$. A random set of points in $\Phi $ can be represented as a countable set of $\lbrace x_{\imath}\rbrace $ random variables that take values in $ \mathbb{R}^2$. The intensity measure of $\Phi $ is $ \Lambda(B)= \mathbb{E}[\Phi(B)]$, where $\mathbb{E}[\Phi(B)]$ is the expected number of points in $ B \subset \mathbb{R}^2$. A simple PP $\Phi$ is determined by its void probabilities over all compact sets, i.e.\ $ \mathbb{P}(\Phi(B)=0)$ for a compact set $B \subset \mathbb{R}^2$. A point process is said to be stationary if its distribution is invariant with respect to translation (shifts in space) \cite{haenggi2009stochastic} \cite{baddeley2007stochastic}. \\
A stationary Poisson point process (PPP) of intensity $ \lambda_p$ is characterized by the following two properties:
\begin{itemize}
\item The number of points in any set $B \subset \mathbb{R}^2$ is a Poisson random variable with mean $\lambda|B| $, i.e.\
 \begin{small}
 \begin{equation}\label{prob_poisson}
 {P}(\Phi(B)=k)=e^{-\lambda|B|}\dfrac{(\lambda|B|)^k}{k!} 
 \end{equation}
 \end{small}
\item The number of points in disjoint sets are independent random variables \cite{serfozo2009basics}. 
\end{itemize}
\vspace{+0.02in}
\textbf{Campbell's Theorem.} Let $ f(x):\mathbb{R}^2\rightarrow[0,\infty]$ be a measurable integrable function. Then, the average sum of a function evaluated at the points of $\Phi$ is given by: 
\vspace{-0.02in}
\begin{small}
\begin{equation}\nonumber
{E}\left[ \sum_{x\in\Phi}f(x)\right] =\int_{\mathbb{R}^2} f(x)\Lambda(\mathrm{d}x) 
\end{equation}
\end{small}
For a stationary PP $\Phi$, the average number of points in a set $ B \subset \mathbb{R}^2$ conditioning on having a point at the origin "o" but excluding that point is denoted as $ \mathbb{E}^{!o}[ \sum_{x\in\Phi}1_B(x)]$, where $1_B(.)$ is the indicator function \cite{haenggi2009interference}. \\
 If $f(x):\mathbb{R}^2\rightarrow[0,\infty]$ is an integrable function, then 
 \begin{small}
  \begin{equation}\label{red_exp}
 {E}^{!o}\left[ \sum_{x\in\Phi}f(x)\right] =\lambda^{-1}\int_{\mathbb{R}^2}\rho^{(2)}(x) f(x) \mathrm{d}x
 \end{equation}
  \end{small}
\noindent where $ \rho^{(2)}(x)$ is the second order product density of the stationary PP $\Phi$. \\
The conditional probability generating functional (PGFL) of a PP $\Phi$ is given by  
 $$ {G}\left[ f(x)\right] =\mathbb{E}^{!o}\left[ \prod_{x\in\Phi}f(x)\right] $$
\subsection{The Matern Point Process}
A Matern Hard-core (MHC) point process $\Phi_m$ is generated by a dependent thinning of a stationary Poisson point process as follows \cite{stoyan1985one}:
\begin{enumerate}
\item Generate a PPP $\Phi_p$ with density $\lambda_p$.
\item For each point $x\in\Phi_p$ associate a  mark $m_x \sim U[0,1]$ independent of any other point.
\item A point $x$ is retained in $\Phi_m$ if it has the lowest mark compared to all points in $B(x,d)$, i.e.\ a circle centered at $x$ with radius $d$.
\end{enumerate}
The probability of an arbitrary point $x$ is retained in $\Phi_m$ is given by:
\begin{equation}\label{p}
p=\dfrac{1-\exp(-\lambda_p \pi d^2)}{\lambda_p\pi d^2}
\end{equation}
The density of the MHC PP $\Phi_m$ is $\lambda_m = p\lambda_p$, i.e.\
\begin{equation}\nonumber
\lambda_m=\dfrac{1-\exp(-\lambda_p \pi d^2)}{\pi d^2}
\end{equation}
The second order product density of the MHC PP $\Phi_m$ is given by \cite{illian2008statistical}
\vspace{-0.1in}
\begin{equation}\label{2_den}
     \rho^{(2)}(\upsilon)= 
\begin{cases}
\lambda_m^2 , & \text{if } \upsilon\geq2d \\ 
     \dfrac{2V(\upsilon)[1-\exp(-\lambda_p \pi d^2)]}{\pi d^2 V(\upsilon) [ V(\upsilon)-\pi d^2 ]} & \text{if } 2d>\upsilon>d\ \\
     -  \dfrac{2\pi d^2[1-\exp(-\lambda_pV(\upsilon))] }{\pi d^2 V(\upsilon) [ V(\upsilon)-\pi d^2 ]}
\\    0,              & \text{otherwise}
\end{cases}
\end{equation}
where $ V(\upsilon) $ is the union area of two discs of radius $d$ and inter-center distance $\upsilon$, centered at any two points of the MHC point process $\Phi_m$, and $ V(\upsilon) $  is defined as 
\begin{equation}\nonumber
V(\upsilon)=2\pi d^2 - 2d^2 \cos^{-1}\left( \dfrac{\upsilon}{2d}\right)+ \upsilon \sqrt{d^2 - \dfrac{\upsilon^2}{4}} 
\end{equation}
\begin{figure}
\includegraphics[scale=0.55]{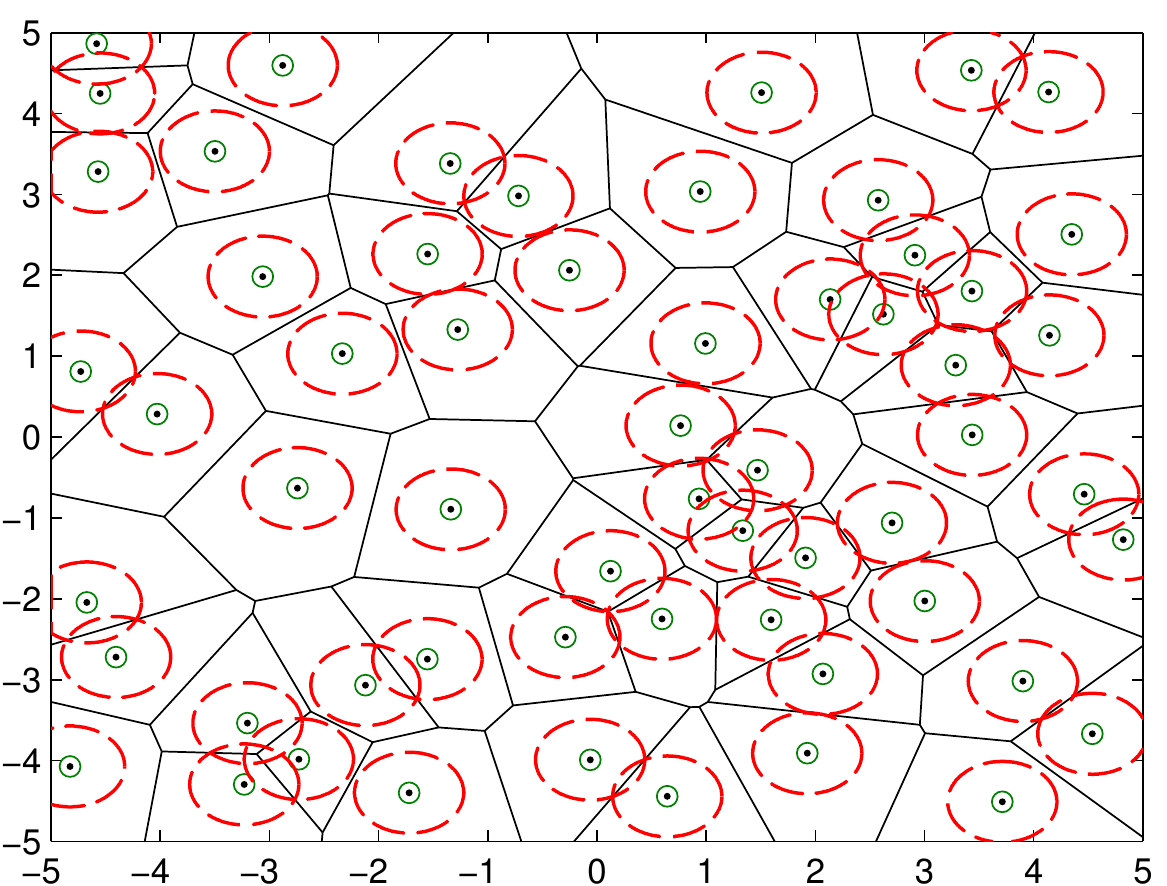}
\centering
\caption{A realization of MHC PP with $\lambda_p=1 , d=0.5$}\label{fig:real}
\end{figure}

\section{System Model}\label{sec:systemmodel}
We model the downlink of a cellular network where the BS deployment is based on a repulsive point process which is a variation of the independent PPP. A repulsive point process guarantees min distance $d$ between BS deployment locations. In this paper, we consider a mathematically tractable type of the repulsive point processes, namely the Matern hardcore (MHC) point process $\Phi_m $ of intensity $\lambda_m $ and a minimum distance $d$ between BSs. We assume that the mobile users spatial distribution follows an independent homogeneous PPP. We assume that a mobile user is connected to the nearest BS. Hence, the base stations downlink coverage areas are Voronoi tessellations on the plane as shown in Fig. \ref{fig:real}.
\begin{figure}
\includegraphics[width=0.35\textwidth, height=0.23\textheight]{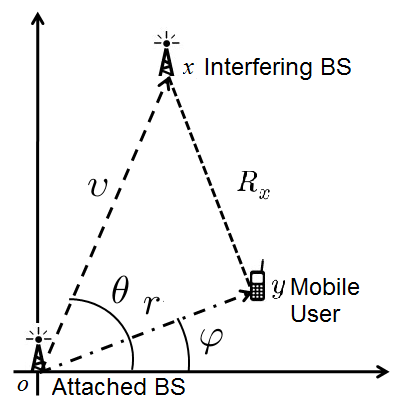}
\centering
\caption{System model}\label{fig:top}
\end{figure}
We adopt the standard path loss propagation model with path loss exponent $\alpha$ and assume that the channel between the mobile user and the attached BS varies according to Rayleigh fading with constant transmitted power $(1/\gamma)$ and noise power $\sigma^2$. Thus, the received power for a typical mobile user at a distance $r$ from the attached BS is $hr^{-\alpha}$, where $h$ is an exponentially distributed random variable, i.e.\ $h\sim\exp(\gamma)$. Moreover, since the MHC PP is a stationary PP we can assume without loss of generality that the attached BS is located at the origin. \\ \indent Our system model is illustrated in Fig. \ref{fig:top}, where $r$ is the distance between the mobile user at point $y$ and the attached BS, $\varphi$ is the angle between the $x$-axis and the vector $r$, $\upsilon$ is the distance between the attached BS and any interfering BS at point $x$ on the plane, $\theta$ is the angle between the $x$-axis and the vector $\upsilon$, and $R_x$ is the distance between the mobile user and any interfering BS at point $x$.
\\ \indent The interference power, $I_r$, is defined as the sum of the received powers from the interfering BSs, i.e.\ other than the attached BS. Thus, the interference power $I_r$ under Rayleigh fading, i.e.\ (an exponentially distributed interference power $g_x\sim\exp(\gamma)$) is defined as
\begin{equation}\nonumber
I_r=\sum_{x\in\phi_m \backslash \lbrace o\rbrace} \!\!\! g_x R_x^{-\alpha}
\end{equation} 
Hence, the SINR for a typical mobile user is defined as follows: \begin{equation}\nonumber
\text{SINR}=\dfrac{hr^{-\alpha}}{\sigma^2 + I_r}
\end{equation}      
\indent The coverage probability is defined as the probability that a typical mobile user is able to achieve some threshold SINR, denoted $\beta$, i.e.\  $ P_c = \mathbb{P}[SINR \geq \beta] $. That is, the probability of coverage is the complementary cumulative distribution function (CCDF) of the SINRs over the network.\\ 

\section{Coverage Analysis}\label{sec:coverage}
Our prime objective in this section is to characterize, analytically, the coverage probability under a MHC spatial point process. 
 Towards this objective, we first derive the MHC empty space distribution in Section \ref{sec:coverage}.A which is an essential step in the derivation of the $P_{c}$. Section \ref{sec:coverage}.B is then dedicated to the major results of this paper presented in Theorem 1 and Proposition 1 which establish lower bounds on the probability of coverage under the MHC spatial point process. 

\subsection{MHC empty space distribution}\label{sec:lemma}
In our model, we assume that a mobile user connects to the nearest BS. Thus, if a mobile user is at a distance $r$ from the attached BS, then there is no interfering BS that is closer than $r$ to the mobile user. The probability density function (pdf) of $r$ is the empty space distribution of the underlying MHC point process which is approximated in the following lemma.\\
\newline\indent \textbf{Lemma 1.}
Given that the mobile user is at a distance $r$ from the attached BS, the approximated MHC empty space distribution $f(r)$ is given by:
\begin{equation}\label{fr}
f(r)= 2\pi\lambda_m r \ {\large e}^{(- \pi \lambda_m r^2)}  
\end{equation}
\vspace{-0.2in}
\begin{proof} See Appendix A.
\end{proof}
\subsection{Probability of Coverage}\label{sec:Thoerems}
This section hosts the main analytical findings of this paper presented in Theorem 1 and Proposition 1. First, we establish a lower bound on $P_{c}$ using Jensen's inequality in Theorem 1. Next, we establish a tighter lower bound on $P_{c}$ in Proposition 1 using the inequality proposed in Conjecture 1. \\
 \newline \indent \textbf{Theorem 1.} A lower bound on the coverage probability for a mobile user in a cellular network deployed using a MHC PP is given by
\begin{equation}\label{theorem1}
P_c\geq \int_{\varphi=0}^{2\pi} \! \int^{\infty}_{r=0} \!\! \dfrac{f(r)}{2\pi} \ e^{-\gamma\beta \sigma^2 r^\alpha}e^{-(\mu_1 + \mu_2)} \mathrm{d}r \ \mathrm{d}\varphi \\
\end{equation}
where
\vspace{-0.1in}
\begin{equation}\label{muo12}
\begin{aligned}
 &\mu_1= \lambda_m^{-1} \int_{\theta=0}^{2\pi} \int_{\upsilon=max[d,|2r\cos(\theta-\varphi)|]}^{max[2d,|2r\cos(\theta-\varphi)|]}  \!\!\!\!\!\!\!\!\!\!\!\!\!\!\!\!\!\!\!\!\!\!\!\!\!\!\!\!\!\!\!\!\!\!\!\!\!\!\Delta(r,\upsilon ,\theta,\varphi) \ \rho_1^{(2)}(\upsilon) \ \upsilon \  \mathrm{d}\upsilon \ \mathrm{d}\theta, \\
 &\mu_2= \lambda_m^{-1} \int_{\theta=0}^{2\pi} \int_{\upsilon=max[2d,|2r\cos(\theta-\varphi)|]}^{\infty} \!\!\!\!\!\!\!\!\!\!\!\!\!\!\!\!\!\!\!\!\!\!\!\!\!\!\!\!\!\!\!\!\!\!\!\!\!\!\!\!\! \Delta(r,\upsilon ,\theta,\varphi) \ \rho_2^{(2)}(\upsilon) \ \upsilon \  \mathrm{d}\upsilon \ \mathrm{d}\theta \\  
\end{aligned}
\end{equation}
and 
\vspace{-0.1in}
$$ \Delta(r,\upsilon,\theta,\varphi)= \ln\left( 1+\beta \left( \dfrac{r^2}{\upsilon^2+r^2-2r\upsilon \cos(\theta-\varphi)}\right)^{\alpha/2} \right) $$
\begin{equation}\label{rho_th1}
\rho^{(2)}(\upsilon)=  
 \begin{cases}
  \rho_1^{(2)}(\upsilon) & \text{,if } d<\upsilon  \leq 2d \\ 
  \rho_2^{(2)}(\upsilon)& \text{,if } \upsilon > 2d \\    
 \end{cases}
\end{equation}
\begin{proof}
See Appendix B.
\end{proof}
\vspace{-0.1in}
Next, we propose an inequality in Conjecture 1 based on our numerical observations, by which we develop a lower bound in Proposition 1 tighter than Theorem 1.\\
 \newline \indent \textbf{Conjecture 1.} For $d$ small enough, the following inequality holds for a MHC PP
\begin{equation}\label{Conjecture1}
\mathbb{E}_{\phi_m}^{!o} \left[ \prod_{x \in \phi_m} 1-\Delta_x \right]  \geq e^{ -\mathbb{E}_{\phi_m}^{!o}\left[ \sum\limits_{x \in \phi_m} \Delta_x\right]} 
\end{equation}
where 
\begin{equation}\label{delta}
\Delta_x=\dfrac{1}{1+\beta^{-1} \left( \frac{r}{R_x}\right)^{-\alpha}}
\end{equation}

The inequality proposed in Conjecture 1 is motivated by the fact that for $d$ small enough, the probability of coverage of a MHC PP $\geq$ the probability of coverage of a PPP and the PGFL of PPP is given by 
{\small $$\mathbb{E}_{\phi_p}^{!o} \left[ \prod\limits_{x \in \phi_p} 1-\Delta_x \right] = \exp \left(  -\mathbb{E}_{\phi_p}^{!o}\left[  \sum\limits_{x \in \phi_p} \Delta_x\right]  \right)$$}
Thus, applying the PPP PGFL definition on a MHC PP results in a lower bound on MHC $P_{c}$ which is characterized by the inequality in Conjecture 1.\\ \\ 
\indent \textbf{Proposition 1.} A lower bound on the coverage probability for a mobile user in a cellular network deployed using a MHC PP is given by
\begin{equation}\nonumber
\begin{aligned}
P_c\geq \int_{\varphi=0}^{2\pi} \! \int^{\infty}_{r=0} \!\! \dfrac{f(r)}{2\pi} \ e^{-\gamma\beta \sigma^2 r^\alpha}e^{-(\mu_1 + \mu_2)} \mathrm{d}r \ \mathrm{d}\varphi  
\end{aligned}
\end{equation}
with \\
{\small $\Delta(r,\upsilon,\theta,\varphi)= \left( 1+\beta^{-1} \left( \dfrac{r^2}{\upsilon^2+r^2-2r\upsilon \cos(\theta-\varphi)}\right)^{-\alpha/2}\right)^{-1}$}  \\
and $\mu_1 ,\mu_2 ,\rho^{(2)}(\upsilon)$ are the same as (\ref{muo12}) and (\ref{rho_th1}).
\begin{proof}
See Appendix D.
\end{proof}
\section{Numerical Results}\label{sec:results}
\begin{figure}
\includegraphics[scale=0.6]{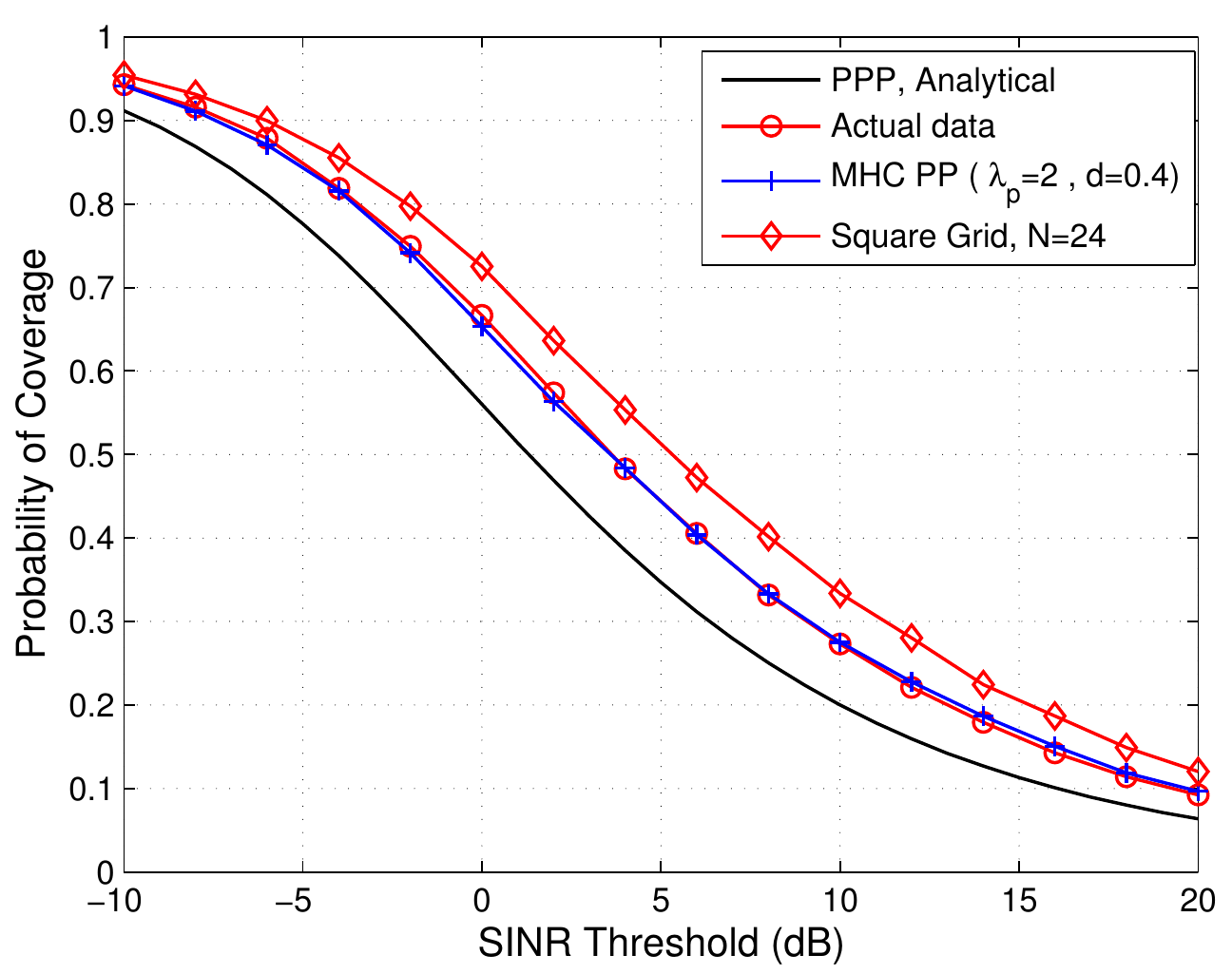}
\centering
\caption{Comparison of the coverage probability for a PPP, a MHC with $\lambda_p=2,d=0.4$, a square grid and actual data}\label{fig:Pc}
\end{figure}

\begin{figure}
\includegraphics[scale=0.6]{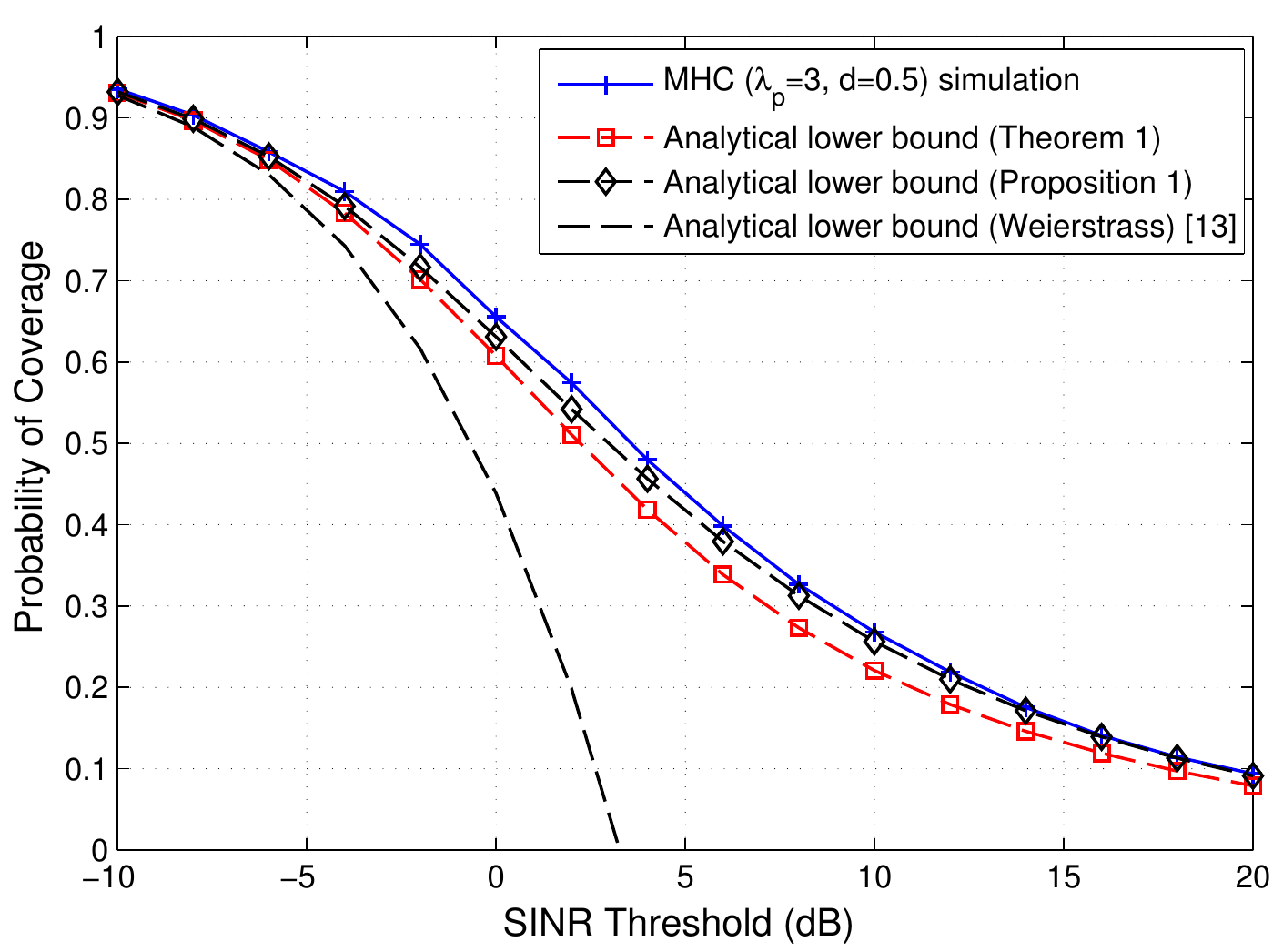}
\centering
\caption{Analytical lower bounds for a MHC scenario with $ \lambda_p=3 $ and $ d=0.5 $ }\label{fig:bounds}
\end{figure}

First, we compare the $P_{c}$ of the MHC PP against a PPP, a square grid and an actual BS deployment from a rural area \cite{sitefinder}. Intuitively, we expect that the $P_{c}$ of the MHC PP to be bounded by the PPP as a lower bound and the grid model as an upper bound. Also, we show that an actual BS deployment can be fitted by choosing the appropriate $\lambda_p$ and $d$ of the MHC PP. Second, we solve the integrals of the analytical lower bounds introduced in Theorem 1 and Proposition 1 numerically and compare them against a simulated MHC scenario.\\
In Fig. \ref{fig:Pc}, we compare the $P_{c}$ for different models under Rayleigh fading and path loss exponent $ \alpha=4 $. We set the noise power to $\sigma^2=0.1P_T$, where $P_T$ is the BS transmitted power. It can be noticed from Fig. \ref{fig:Pc} that the $P_{c}$ derived analytically for the PPP in \cite{andrews2011tractable} under Rayleigh fading yields the most conservative $P_{c}$ and the square grid lattice with 24 BSs yields the most optimistic $P_{c}$ which agrees with intuition.
Therefore, the $P_{c}$ of the MHC PP and the $P_{c}$ of an actual BS deployment in $100\times80$ km rural area lie between the $P_{c}$ of the PPP and the grid models, assuming all BSs are omni-directional and transmit with unit power. The MHC PP parameters $\lambda_p=2$ and $d=0.4$ are tuned to fit the actual data.   
\indent In Fig .\ref{fig:bounds}, we solve the lower bounds integrals numerically using the composite trapezoidal rule and compare them against a simulated MHC scenario with $\lambda_p=3$ and $d=0.5$. It can be noticed that the lower bound introduced in Proposition 1 is tighter than Theorem 1 within $4\%$ from the simulated data on the average and it is quite accurate in plausible scenarios where the SINR threshold ranges from $10$ to $20$ dB. Finally, we replace Conjecture 1 by the Weierstrass inequality in \cite{ganti2010new} and we notice that the $P_{c}$ diverges significantly as we increase the SINR threshold. 
  \begin{figure}
\includegraphics[width=0.45\textwidth, height=0.25\textheight]{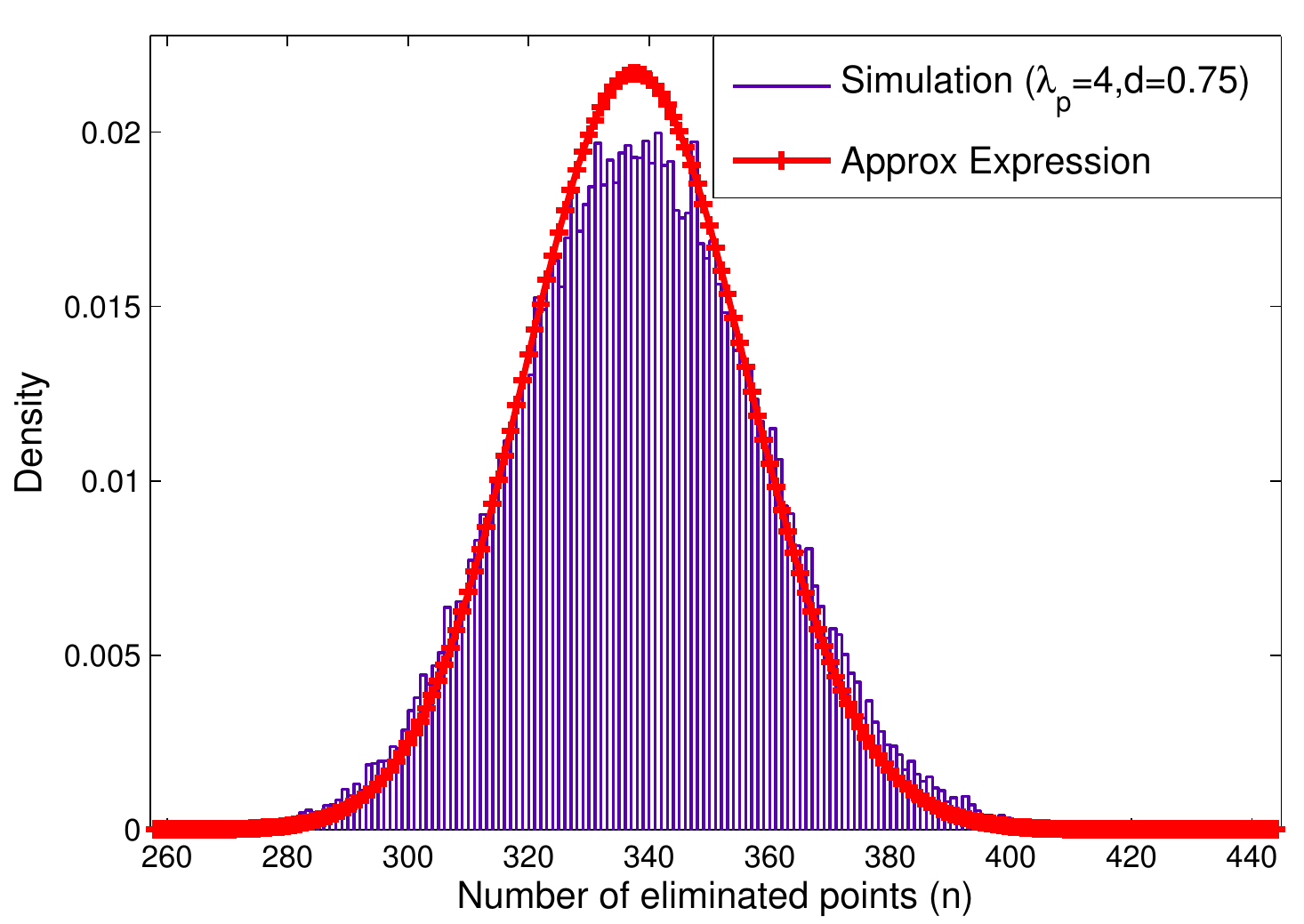}
\centering
\caption{Probability distribution of $n$ eliminated points}\label{fig:npoints}
\end{figure}

\section{Conclusions}\label{sec:conclusion}
We presented a stochastic geometry formulation using the Matern hardcore spatial point process to model the BS deployment in cellular networks. Nevertheless, the presented analysis can be employed to study other wireless networks, e.g.\ ad hoc networks. This paper constitutes a departure from the recent literature on studying and analyzing coverage probability using independent Poisson point processes. We established, analytically, two lower bounds for the coverage probability which constitutes our major analytical findings and constitutes an important step towards deriving closed form expressions. We compared our model to actual BSs locations from a rural area and showed that the actual data can by fitted by using the appropriate MHC density $\lambda_m$ via appropriately tuning $\lambda_p \text{ and } d$. An important future extension of this work is to build on our analytical findings to characterize the MHC $P_c$ in closed-form. Other future directions could be modeling multi-tier cellular networks and   incorporating other repulsive point processes.
\section*{Appendix}\label{sec:appendeix}
\subsection{Proof of Lemma.1}
\vspace{-0.2 in}
$$ \mathbb{P}[\text{No BS is closer than $r$}]= \mathbb{P}[N_{{\tiny M}}(B(y,r))=0)]$$
, where $N_{{\tiny M}}(B(y,r)) $ is the number of points in a circle $B(y,r)$ centered at the user $y$ with radius $r$ in a MHC PP.
 The $\mathbb{P}[N_{{\tiny M}}(B(y,r))=0)] $ in a MHC PP is equal to 
 $$\sum_{n=0}^{\infty} \mathbb{P}[N_{{\tiny P}}(B(y,r))=n]\mathbb{P}[\text{$n$ points are eliminated in MHC}]$$
 , where $N_{{\tiny P}}(B(y,r)) $ is the number of points in $B(y,r)$ from the original PPP. The probability that a point is eliminated in the MHC PP is equal to $q=1-p$, where $p$ is given by (\ref{p}).
 Using (\ref{prob_poisson}), we get
 {\small $$ \mathbb{P}[N_{{\tiny P}}(B(y,r))=n]= \sum_{n=0}^{\infty}\exp(-\lambda_p\pi r^2)\dfrac{(\lambda_p\pi r^2)^n}{n!}$$}
Then, we have the following approximation: \\ 
 For $d$ small enough,\\ $\mathbb{P}[\text{$n$ points are eliminated}]\approx(1-p)^n=(1-(\frac{\lambda_m}{\lambda_p}))^n $ , hence
 \begin{equation}\label{case_a}
 \begin{aligned}
 \mathbb{P}[N_{{\tiny M}}(B(y,r))=0] &=\sum_{n=0}^{\infty}\exp(-\lambda_p\pi r^2)\dfrac{((\lambda_p-\lambda_m)\pi r^2)^n}{n!}
 \\ &=\exp(-\lambda_m \pi r^2)
 \end{aligned}
 \end{equation} 
  \\ Applying $f(r)= \dfrac{\mathrm{d}}{\mathrm{d}r} \left( 1-\mathbb{P}[N_{{\tiny M}}(B(y,r))=0]\right) $ to (\ref{case_a}) we get (\ref{fr}). \\
 The approximation in (\ref{case_a}) is shown in Fig. \ref{fig:npoints}. Also, we have checked numerically that the approximation in (\ref{fr}) does not affect the $P_c$ results.
 
\subsection{Proof of Theorem 1}
The probability of coverage ($P_{c}$) is equal to 
\begin{equation}\nonumber
\begin{aligned}
P_c&=\mathbb{E}_{ r,\varphi}[\mathbb{P}(\text{SINR}>\beta | r,\varphi)] \\
&=\int_{\varphi=0}^{2\pi} \! \int^{\infty}_{r=0} \!\! \dfrac{f(r)}{2\pi} \ \mathbb{P}(\text{SINR}>\beta | r,\varphi) \ \mathrm{d}r  \mathrm{d}\varphi \\
\end{aligned}
\end{equation}
\begin{equation}\nonumber
\begin{aligned}
\mathbb{P}(\text{SINR}>\beta | r,\varphi)&=\mathbb{P}(h >\beta(\sigma^2 + I_r)r^{\alpha} | r,\varphi) \\
&\stackrel{(a)}{=}\mathbb{E}^{!o}_{I_r}\left[e^{-\gamma(\beta(\sigma^2 + I_r)r^{\alpha})} | r,\varphi \right]   
\end{aligned}
\end{equation}
step (a) assuming Rayleigh fading, i.e.\ $h\sim\exp(\gamma)$
\begin{equation}\label{Pc_general}
\begin{aligned}
P_c=\int_{\varphi=0}^{2\pi} \! \int^{\infty}_{r=0} \!\! \dfrac{f(r)}{2\pi} \ e^{-\gamma\beta \sigma^2r^{\alpha}} \ \mathbb{E}^{!o}_{I_r}\left[e^{-\beta \gamma r^{\alpha} I_r} |r,\varphi \right]  \mathrm{d}r  \mathrm{d}\varphi 
\end{aligned}
\end{equation}
where 
\begin{equation}\label{eqn:laplace}
\begin{aligned}
\mathbb{E}^{!o}_{I_r}\left[e^{-\beta \gamma r^{\alpha} I_r} |r,\varphi \right]&= \mathbb{E}_{\phi_m,g_x}^{!o}\left[e^{-\beta \gamma r^{\alpha} \!\!\!\! \sum\limits_{x\in\phi_m} \!\!\! g_x R_x^{-\alpha}} \right] \\
&\stackrel{(a)}{=}\mathbb{E}_{\phi_m}^{!o}\left[ \prod\limits_{x \in \phi_m} \mathbb{E}_{g_x} \left[ e^{-\beta \gamma r^{\alpha}  g_x R_x^{-\alpha}}\right]  \right] \\
&\stackrel{(b)}{=} \mathbb{E}_{\phi_m}^{!o}\left[ \prod\limits_{x \in \phi_m}  \dfrac{1}{1+\beta \left( \frac{r}{R_x}\right)^{\alpha}} \right] \\
&=\mathbb{E}_{\phi_m}^{!o}\left[ e^{ - \sum\limits_{x \in \phi_m} \ln\left( 1+\beta \left( \frac{r}{R_x}\right)^{\alpha}\right) }\right] \\
&\stackrel{(c)}{\geq}  e^{\mathbb{E}_{\phi_m}^{!o}\left[ - \sum\limits_{x \in \phi_m} \ln\left( 1+\beta \left( \frac{r}{R_x}\right)^{\alpha}\right) \right]} \\
\end{aligned}
\end{equation}
Steps (b) from $g_x\sim\exp(\gamma)$ and step (c) using Jensen's inequality, let 
\begin{equation}\nonumber
\begin{aligned}
\mu=\mathbb{E}_{\phi_m}^{!o}\left[ - \sum\limits_{x \in \phi_m} \Delta_x \right] 
\text{, where } \Delta_x =\ln\left( 1+\beta \left( \frac{r}{R_x}\right)^{\alpha}\right) \\
\end{aligned}
\end{equation} 
hence, 
\begin{equation}\nonumber
\begin{aligned}
P_c\geq\int_{\varphi=0}^{2\pi} \! \int^{\infty}_{r=0} \!\! \dfrac{f(r)}{2\pi} \ e^{-\gamma\beta \sigma^2r^{\alpha}} \ e^{-\mu} \ \mathrm{d}r  \mathrm{d}\varphi 
\end{aligned}
\end{equation} 
From trigonometry of Fig .\ref{fig:top}, $R_x$ can be substituted by 
$$R_x=\sqrt{\upsilon^2 + r^2 -2r\upsilon \cos(\theta-\varphi)} $$ hence, $ \Delta(r,\upsilon,\theta,\varphi)=\ln\left( 1+\beta \left( \frac{r^2}{\upsilon^2 + r^2 -2r\upsilon \cos(\theta-\varphi)}\right)^{\alpha/2}\right)$
From (\ref{red_exp}) , $\mu$ is equal to 
\begin{equation}\nonumber
\begin{aligned}
\mu=\lambda_m^{-1}\int_{\theta=0}^{2\pi} \int_{\upsilon=max[d,|2r\cos(\theta-\varphi)|]}^{\infty} \!\!\!\!\!\!\!\!\!\!\!\!\!\!\!\!\!\!\!\!\!\!\!\!\!\!\!\!\!\!\!\!\!\!\!\!\!\!\!\!\!\!\!\!\!\!\ \rho^{(2)}(\upsilon) \ \Delta(r,\upsilon,\theta,\varphi) \ \upsilon  \mathrm{d}\upsilon\mathrm{d}\theta
\end{aligned}
\end{equation} 
Using the definition of $\rho^{(2)}(\upsilon)$ in (\ref{2_den}), then $\mu=\mu_1+\mu_2$, where
\begin{equation}\nonumber
\mu_1 \stackrel{(a)}{=} \lambda_m^{-1} \int_{\theta=0}^{2\pi} \int_{\upsilon=max[d,|2r\cos(\theta-\varphi)|]}^{max[2d,|2r\cos(\theta-\varphi)|]} \!\!\!\!\!\!\!\!\!\!\!\!\!\!\!\!\!\!\!\!\!\!\!\!\!\!\!\!\!\!\!\!\!\!\!\!\!\!\Delta(r,\upsilon,\theta,\varphi) \ \rho_1^{(2)}(\upsilon) \ \upsilon \  \mathrm{d}\upsilon \ \mathrm{d}\theta \\
\end{equation}
\begin{equation}\nonumber
\mu_2 \stackrel{(b)}{=} \lambda_m^{-1} \int_{\theta=0}^{2\pi} \int_{\upsilon=max[2d,|2r\cos(\theta-\varphi)|]}^{\infty} \!\!\!\!\!\!\!\!\!\!\!\!\!\!\!\!\!\!\!\!\!\!\!\!\!\!\!\!\!\!\!\!\!\!\!\!\!\!\!\!\!\!\!\!\!\!\Delta(r,\upsilon,\theta,\varphi) \ \rho_2^{(2)}(\upsilon) \ \upsilon \  \mathrm{d}\upsilon \ \mathrm{d}\theta \\ 
\end{equation}
\begin{equation}\nonumber
\rho^{(2)}(\upsilon)=  
 \begin{cases}
  \rho_1^{(2)}(\upsilon) & \text{,if } d<\upsilon  \leq 2d \\ 
  \rho_2^{(2)}(\upsilon)& \text{,if } \upsilon > 2d \\
  \end{cases}
\end{equation} 
From trigonometry of Fig. \ref{fig:top}, $ \upsilon=2r\cos(\theta-\varphi)$ for $R_x=r$. 
Thus, the integral limits of $ \upsilon$ is from $max[d,|2r\cos(\theta-\varphi)|]$ to $\infty $, since the closest interfering BS is at least at distance $r$ and $\rho^{(2)}(\upsilon)=0$ for $\upsilon<d$.

\subsection{Proof of Proposition 1}
Same as proof of Theorem 1 till (\ref{eqn:laplace}-b), then proceed with  \\
\begin{small}
\begin{equation}\nonumber
\begin{aligned}
\mathbb{E}_{\phi_m}^{!o}\left[ \prod\limits_{x \in \phi_m}  \dfrac{1}{1+\beta \left( \frac{r}{R_x}\right)^{\alpha}} \right]&=
\mathbb{E}_{\phi_m}^{!o}\left[ \prod\limits_{x \in \phi_m}  1-\Delta_x \right]\\
&\stackrel{(a)}{\geq} \exp\left( -\mathbb{E}_{\phi_m}^{!o}\left[ \!\sum_{x \in \phi_m} \!\!\Delta_x\right] \right) 
\end{aligned}
\end{equation}
\end{small}
step (a) by using Conjecture 1. 

\bibliographystyle{IEEEtran}
\bibliography{IEEEabrv,DiversityLib}


\end{document}